\documentclass[12pt]{article}
\usepackage{epsfig}
\usepackage{amsmath}
\usepackage[psamsfonts]{amssymb}
\usepackage{amsthm}
\usepackage{euscript}

\usepackage{latexsym}

\renewcommand{\theequation}{\arabic{section}.\arabic{equation}}
\renewcommand{\(}{\begin{equation}}
\renewcommand{\)}{end{equation} \vspace{-.05in}\linebreak}

\newcounter{saveeqn}
\newcounter{savealpheqn}

\newcommand{\alpheqn}{\setcounter{saveeqn}{\value{equation}}%
 \stepcounter{saveeqn}\setcounter{equation}{0}%
 \renewcommand{\theequation}{\mbox{\arabic{section}.\arabic{saveeqn}\alph{equation}}}
 \renewcommand{\)}{\end{equation}}}
\def\part#1{\frac{\partial}{\partial{#1}}}%
\def\group#1{\refstepcounter{equation}\setcounter{saveeqn}{\value{equation}}%
 \label{#1}\setcounter{equation}{0}%
 \renewcommand{\theequation}{\mbox{\arabic{section}.\arabic{saveeqn}\alph{equation}}}
 \renewcommand{\)}{\end{equation}}}
\newcommand{\reseteqn}{\setcounter{equation}{\value{saveeqn}}%
 \renewcommand{\theequation}{\arabic{section}.\arabic{equation}}%
 \renewcommand{\)}{\end{equation}}}
\newcounter{alphcount}

\def\writeletter#1{\renewcommand{\theequation}{\alph{#1}}%
                 \begin{eqnarray}%
                 \label{#1}%
                 \nonumber\end{eqnarray}\vspace{-.666in}}
\def\getlette2r#1{\newcounter{#1}%
                 \setcounter{#1}{\value{equation}}%
                 \providecommand{\writeletters}{\writeletters\writeletter{#1}}}

\newcommand{\aalpheqn}{\setcounter{saveeqn}{\value{equation}}%
 \stepcounter{saveeqn}\setcounter{equation}{0}%
 \renewcommand{\theequation}{\mbox{\Alph{subsection}.\arabic{saveeqn}\alph{equation}}}
  \renewcommand{\)}{\end{equation}}}
\newcommand{\areseteqn}{\setcounter{equation}{\value{saveeqn}}%
 \renewcommand{\theequation}{\Alph{subsection}.\arabic{equation}}%
 \renewcommand{\)}{\end{equation}}}

\renewcommand{\thefootnote}{\alph{footnote}}
\renewcommand{\(}{\begin{equation}}
\renewcommand{\)}{\end{equation}}
\newcommand{\ba}{\begin{eqnarray}}
\newcommand{\ea}{\end{eqnarray}}

\newcommand{\bp}{\mathop{\vtop{\ialign{##\crcr
  $\hfil\displaystyle{}\hfil$\crcr\noalign{\kern-13pt\nointerlineskip}
  \BIG{(}\hskip0pt\crcr\noalign{\kern3pt}}}}}
\newcommand{\cbp}{\mathop{\vtop{\ialign{##\crcr
  $\hfil\displaystyle{}\hfil$\crcr\noalign{\kern-13pt\nointerlineskip}
  \BIG{)}\hskip0pt\crcr\noalign{\kern3pt}}}}}
\newcommand{\pa}{\mathop{\vtop{\ialign{##\crcr
  $\hfil\displaystyle{\oplus}\hfil$\crcr\noalign{\kern+1pt\nointerlineskip}
  \hspace{.08in}$^{\alpha=0}$\hskip6pt\crcr\noalign{\kern3pt}}}}}
\renewcommand{\sp}{,\hspace{.3in}}
\newcommand{\newsection}{\setcounter{equation}{0}\section}

\newcommand{\Z}{\ensuremath{\mathbb Z}}

\newcommand{\beq}{\begin{equation}}
\newcommand{\eeq}{\end{equation}}

\def\BIG#1{\mbox{\Huge $#1$}}

\mathchardef\endbar="375 \font\fivesans=cmss10 at 4.61pt
\font\sevensans=cmss10 at 6.81pt
\font\tensans=cmss10 at 12pt 
\newfam\sansfam
\textfont\sansfam=\tensans\scriptfont\sansfam=\sevensans\scriptscriptfont
\sansfam=\fivesans
\def\sans{\fam\sansfam\tensans}
\def\Z{{\mathchoice
{\hbox{$\sans\textstyle Z\kern-0.455em Z$}} 
{\hbox{$\sans\textstyle Z\kern-0.455em Z$}} 
{\hbox{$\sans\scriptstyle Z\kern-0.355em Z$}} 
{\hbox{$\sans\scriptscriptstyle Z\kern-0.255em Z$}}}} 

\font\tensans=cmss10 at 14pt
\newfam\sansfamb
\textfont\sansfamb=\tensans\scriptfont\sansfamb=\sevensans\scriptscriptfont
\sansfamb=\fivesans

\font\tensans=cmss10 at 17pt
\newfam\sansfamc
\textfont\sansfamc=\tensans\scriptfont\sansfamc=\sevensans\scriptscriptfont
\sansfamc=\fivesans

\def\contr#1#2{\mathop{\vtop{\ialign{##\crcr
  $\hfil\displaystyle{#2}\hfil$\crcr\noalign{\kern3pt\nointerlineskip}
  \hspace{.09in}\rule[0in]{.01in}{.1in}\rule[0in]{#1in}{.01in}\rule[0in]{.01in}{.1in}\hskip6pt\crcr\noalign{\kern3pt}}}}}
\def\contrb#1#2#3{\mathop{\vtop{\ialign{##\crcr
  $\hfil\displaystyle{#3}\hfil$\crcr\noalign{\kern3pt\nointerlineskip}
  \hspace{#1in}\rule[0in]{.01in}{.1in}\rule[0in]{#2in}{.01in}\rule[0in]{.01in}{.1in}\hskip6pt\crcr\noalign{\kern3pt}}}}}

\def\hsp#1{\hspace{#1in}}

\catcode`\@=11
\def\vereq#1#2{\lower3pt\vbox{\baselineskip1.5pt \lineskip1.5pt
\ialign{$\m@th#1\hfill##\hfil$\crcr#2\crcr\sim\crcr}}}
\catcode`\@=12

\begin{document}
\begin{titlepage}
\begin{center}

                                        \hfill hep-th/0207089\\

\vskip 1in
\def\thefootnote{\fnsymbol{footnote}}

{\large \bf Spacelike and Time Dependent Branes from DBI\\}

\vskip 0.3in

John E. Wang \footnote{hllywd2@phys.ntu.edu.tw}

\vskip 0.15in

{\em Institute of Physics\\
  Academia Sinica\\
   Taipei, Taiwan }\\

\end{center}


\vfill

\begin{abstract}

Spacelike branes are new time-dependent systems to explore and it
has been observed that related supergravity solutions can be
obtained by analytically continuing known D-brane solutions. Here
we show that analytic continuation of known solutions of the
Dirac-Born-Infeld equations also lead to interesting analogs of
time dependent gravity solutions. Properties of these new
solutions, which are similar to the Witten bubble of nothing and
S-branes, are discussed.  We comment on how these new bubble
solutions seem relevant to the tachyon condensation process of
non-BPS branes, and remark on their application to cosmological
scenarios. Unstable brane configurations which resemble S-brane
type solutions are also discussed.

\end{abstract}

\vfill

\end{titlepage}
\setcounter{footnote}{0}
\renewcommand{\thefootnote}{\arabic{footnote}}

\pagebreak
\renewcommand{\thepage}{\arabic{page}}
\pagebreak
\newsection{Introduction}

Time dependent backgrounds have arisen in the study of tachyon
condensation \cite{SenRollingTachyon, SenTachyonMatter}, as well
as gravitational backgrounds related to de Sitter space
\cite{HullTimeTdual, HullldeSitterMtheory} and inflationary
scenarios \cite{TyeBraneInteractInflation, DvaliTyeBraneInflation,
DvaliShafiDbraneInflation, StephonDDbarInflation,
BurgessInflationBraneaBraneUniverse, GibbonsRollTachyon}. A class
of time dependent solutions called S-branes
\cite{PopeandthreeCosmologicalStringTheory, GutperleStrominger}
has been found, and related supergravity
\cite{PopeandtwoBranesandCosmology, ChenGutperle, MyersSbrane,
DegerKayaSbrane} solutions have already been explored.  In this
paper we approach the topic of time dependent systems by
examining solutions of the Dirac-Born-Infeld action. We find
similar braney objects in the sense that the induced metric on
these solutions is similar to the metric of the known gravity
solutions.

In Section~\ref{ReviewSec} we give a quick review of the DBI
action and its string-like excitation solution \cite{CallanM,
GibbonsBion}.  We discuss how this static brane and anti-brane
configuration resembles charged wormhole solutions from black hole
theory. In Section~\ref{ACSec} we discuss new DBI solutions which
come from analytically continuing the string-like excitation
solutions. These brane configurations are similar to Witten
bubbles\cite{Wittenbubble} and play a role in brane and
anti-brane annihilation.  An important difference between the new
bubble solutions presented here and the previous gravity solution,
is that a Witten bubble ``eats up'' spacetime while the bubble
solution in this paper is a safer process which does not effect
the bulk of spacetime.  In our case we have a time dependent
brane and anti-brane configuration in flat space.

Section~\ref{SbraneSec} introduces solutions which appear similar
to S-brane solutions and which also arise from analytically
continuing the string-like excitation solution. These solutions
are presented and discussed although their interpretation is more
difficult since they are embedded in spaces with two times.

\newsection{Review of DBI} \label{ReviewSec}
In this section we will give a quick review of the DBI action and
its equations of motion.  We review the string-like excitation
\cite{CallanM, GibbonsBion} of a brane and give an explicit
embedding of the solution.  In addition we give the induced metric
on this brane and anti-brane configuration which is a type of
static wormhole.  Discussion of the various p-branes solutions is
separated into three categories $p>3$, $p=3$ and $p<3$ which
exhibit qualitatively different behaviors.

\subsection{Equations of Motion}

The Dirac-Born-Infeld action
\begin{equation}
 S =T_p \int d^{p+1}x \sqrt{-det(G_{MN} \partial_a Z^M
\partial_b Z^N + F_{ab} ) }
\end{equation}
is the worldvolume action describing the shape and abelian gauge
field of a Dp-brane embedded in a $D+1$ dimensional space, $M_2$.
The metric on $M_2$ is $G_{MN}$ and the indices $M,N$ range from
0 to D. The coordinates $x^a, \ a=0,...,p$ are the coordinates
for the brane worldvolume while the functions
\begin{equation}
Z^M : \ M_1 \rightarrow M_2 \ \ \ \forall M
\end{equation}
are the embedding coordinates of the worldvolume manifold $M_1$
into the embedding space $M_2$.  For simplicity we will restrict
ourselves to the case where the embedding space is flat
pseudo-Euclidean spacetime $M_2=R^{(p,q)}$.  We will also take
the Cartesian coordinate representation of $M_2$ so the
Christoffel symbols vanish.

This action provides the low energy effective description of open
string dynamics provided we take the usual light brane or
decoupling limit \cite{Giveonreview} which neglects backreaction
effects of the brane on the target space. The shape of the brane
is obtained by varying the action
\begin{equation}
\delta_Z S= -\frac{T_p}{2}
\int d^{p+1} x \ {\sqrt{-(g+F) }} (g+F)^{ab} \delta ( G_{MN}
\partial_a Z^M \partial_b Z^N) =0 \ .
\end{equation}
The equations of motion for the DBI action are
\begin{eqnarray}
\partial_a ( {\sqrt{-(g+F)}} (g+F)^{ab}_S \partial_b Z^M) & = & 0 \ \ \ \ \forall \ M  \\
\partial_b ( {\sqrt{-(g+F)}} (g+F)^{ba}_A )  & = & 0 \ \ \ \ \forall \ a
\end{eqnarray}
where the second equation comes from varying the action with
respect to the gauge potential. In the above expressions
$(g+F)^{ab}_{A/S}$ is the (anti-)symmetric part of the inverse of
the matrix $g+F$ which is generally different from the sum of
inverses.  It is known that this system of equations can also be
obtained from a pure Born-Infeld system with just a field
strength.

\subsection{String-like Brane Intersections}

\begin{figure}[htb]
\begin{center}
\epsfxsize=4in\leavevmode\epsfbox{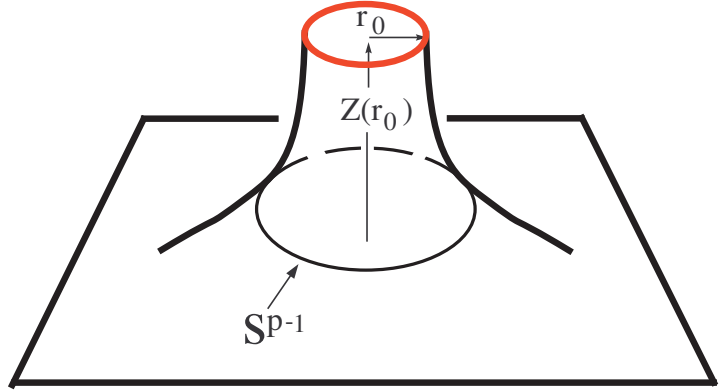} \caption{Static
volcano solution of a Dp brane.} \label{volcano}
\end{center}
\end{figure}

The known string type excitation of the DBI action can be
presented in the following parametrization
\begin{eqnarray} \label{stringembedding}
Z^0 & = & t \\
Z^1 & = & r cos \theta \\
Z^i & = & r sin \theta n^i \sp i=2,...,p\\
Z^{p+1} & = & \int \frac{B}{\sqrt{r^{2p-2} - r_0^{2p-2}}} dr \ .
\end{eqnarray}
The rest of the $Z$ coordinates are constant functions which will
not play a dynamical role in the discussion and can be set to
zero. The above parametrization satisfies the constraints
$\sum_{a=1}^p (Z^a)^2= r^2$ and $\sum_i (n^i)^2=1$. The vector
$\vec{n}$ is a coordinate free way to specify the angular
coordinates of the brane worldvolume; an example of an explicit
parametrization is $n^i=(cos\phi_1, sin\phi_1 cos\phi_2, sin\phi_1
sin \phi_2 cos \phi_3,...)$. The first $p+1$ coordinates are
coordinates of the brane worldvolume and the $Z^{p+1}$ coordinate
is the one dimensional string-like excitation direction of the
brane. For the cases $p \geq 3$ the above solution asymptotes to
$Z^{p+1}=0$ as the radius $r$ approaches infinity.  Also, even
though the derivative $\partial_r Z^{p+1}$ goes to infinity at
$r=r_0$, the value of $Z^{p+1}$ approaches a finite value
$Z^{p+1}(r_0)$ when $r_0 \neq 0$.   Examining this string-like
excitation of the brane, illustrated in Figure~\ref{volcano}, we
find that it has the appearance of a volcano where the radius of
the volcano's mouth is $r_0$ and the height of the volcano is
$Z^{p+1}(r_0)$.

The embedding space is Minkowski with metric
$G_{MN}=\eta_{MN}=(-1,1,1,...,1)$ where $M,N=0,...,D$ so the
induced metric on this brane is
\begin{equation}
ds^2_{volcano} = \eta_{\mu \nu} dZ^\mu dZ^\nu =
-dt^2+\frac{1+A^2/r^{2p-2}}{1-(r_0/r)^{2p-2}} dr^2 + r^2
(d\theta^2 + sin^2\theta d\Omega_{p-2}^2)
\end{equation}
\begin{equation}
A^2 \equiv B^2 - r_0^{2p-2} \ .
\end{equation}
The variable $A$ plays a role in determining the field strength
of this solution
\begin{equation} \label{radialelectric}
F_{tr} = \frac{A}{\sqrt{r^{2p-2} - r_0^{2p-2}}}
\end{equation}
which will play an interesting role later in our discussion. To
insure that the electric field is real valued, the parameters must
satisfy $B^2\geq r_0^{2p-2}$.  We note here that the field
strength diverges at $r=r_0$ and there is no source of charge
along the brane.

\begin{figure}[htb]
\begin{center} \epsfxsize=3in\leavevmode\epsfbox{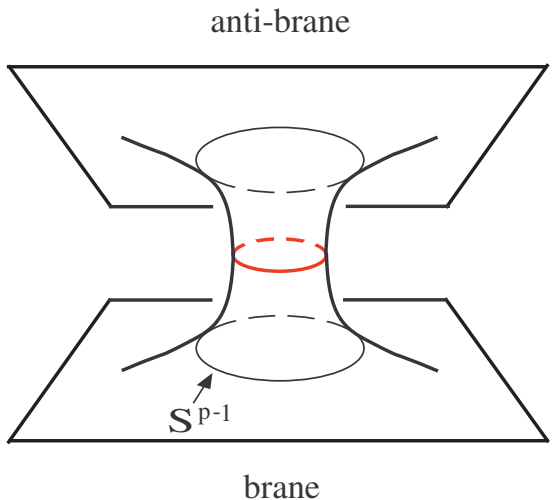}
\caption{Two volcanos, representing a brane and anti-brane, are
joined to form a wormhole.} \label{wormhole}
\end{center}
\end{figure}

If we glue two such volcano solutions together at the radius
$r=r_0$ then this is a static configuration, shown in
Figure~\ref{wormhole}, representing a brane and anti-brane
configuration connected by a wormhole throat.  An explicit way to
do this is to take two copies of the above parametrization but
with
\begin{equation}
Z^{p+1}=2 Z^{p+1}(r_0) - \int \frac{B}{\sqrt{r^{2p-2} -
r_0^{2p-2}}} dr
\end{equation}
for the second copy. We note that this configuration is similar
to the Einstein-Rosen bridge where the radius $r_0$ corresponds
to the horizon and the brane and the anti-brane correspond to the
two asymptotically Minkowski regions.  This solution is smooth
and does not have a singular region of infinite curvature.  It is
like a charged black hole where the constant $A$ is an electric
charge parameter, $B$ the mass parameter and $r_0$ tells us how
far away we are from the BPS configuration.

Several limits are of interest. The first is the limit, $A^2=B^2$
where the electric charge is equal to the mass.  In this limit
$r_0$ is zero and the string excitation stretches to infinity.
This BPS configuration is interpreted as a half-infinite string
ending on a brane with the end of the open string acting as an
electric charge source for the gauge field on the brane. There is
also the interesting non-BPS limit $A=0$ with no electric
charge.  This solution
\begin{equation}
ds^2_{no \ charge \ volcano} = -dt^2+\frac{dr^2}{1-(r_0/r)^{2p-2}}
+ r^2 (d\theta^2 + sin^2\theta d\Omega_{p-2}^2)
\end{equation}
is similar to a Schwarzschild black hole. However whereas the
Schwarzschild black hole can be extended past the horizon, the
above parametrization does not reach into the singular region.
Also the time direction of this volcano solution is independent
of the radius, so dynamics on this brane will not be the same as
on a Schwarzschild spacetime \cite{CallanM}.

Because the volcano solutions are not supersymmetric they have
also been mentioned as being related to possible decay modes for
brane and anti-brane annihilation.  However since these solutions
are time independent, it is interesting to obtain explicit time
dependent annihilation processes. In the next section we will
discuss new time dependent solutions and their relevance to brane
and anti-brane annihilation.

\subsection{Low Dimensional Volcano Solutions} \label{lowdimvolcanosec}
 For the case of $D3$ branes, there is a
similar configuration to the one mentioned above where the role
of electric and magnetic charges is switched.  In this case we
turn on a magnetic field in the radial direction
\begin{equation}  \label{volcanomagnets}
F_{\theta \phi}= \sqrt{B^2-r_0^4}sin \theta \ \ \Leftrightarrow \
\ B_r=\sqrt{\frac{1-r_0^4/r^4}{1+A^2/r^4}}\frac{A}{r^2}
\end{equation}
in which case the infinite excitation is not a fundamental string
but a D-string.

\begin{figure}[htb]
\begin{center}
\epsfxsize=2in\leavevmode\epsfbox{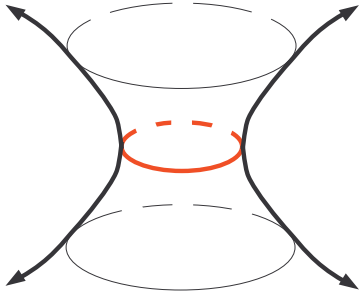} \caption{The
solution for two branes is logarithmic and does not asymptote to
a fixed value.} \label{D2brane}
\end{center}
\end{figure}

For the case of two-branes, the excitation does not asymptote to
a fixed value in the large radius $r$ limit.  The integral
expression for $Z^{3}$ can be solved explicitly in this case and
has logarithmic behaviour
\begin{equation}
r \rightarrow \infty \ \ \Rightarrow \ \ Z^{3}=B \
\textrm{ln}\frac{1}{2} | \sqrt{r^2} + \sqrt{r^2-r_0^2} |
\rightarrow \ \textrm{ln} \ r \ .
\end{equation}
This logarithmic behavior, shown in Figure~\ref{D2brane}, is
expected for a system of two-branes instead of two-branes
connected by a string, and indicates that we are capturing
non-perturbative string coupling effects.

\begin{figure}[htb]
\begin{center}
\epsfxsize=2.5in\leavevmode\epsfbox{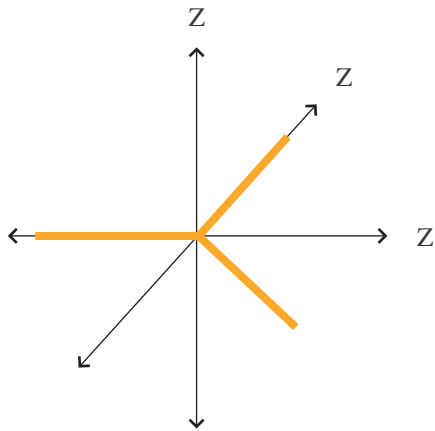} \caption{Three
string junction with spatial directions labeled schematically by
$Z$.} \label{junction}
\end{center}
\end{figure}

Finally, for the case $p=1$ we obtain string junction
configurations which are continuous but not necessarily smooth.
An example of a three string junction is shown in
Figure~\ref{junction}.

\newsection{Bubble Solution} \label{ACSec}

Given a solution to the DBI equations, we can generate new
solutions by analytic continuation. Here we will use this
procedure to generate time dependent solutions which seem to play
a role in brane and anti-brane annihilation. Specifically we
analytically continue the volcano solution of the previous section
and obtain an analogue of Witten's bubble of nothing.  This is a
DBI approach to brane and anti-brane annihilation with some
similarities to the supergravity approach discussed by Fabinger
and Horava \cite{FabingerHorava}. They studied a
non-supersymmetric $E_8 \times \overline{E_8}$ compactification
of M theory on $S_1/Z_2$ and proposed that spacetime could
semiclassically annihilate due to wormholes nucleating between
the two $E_8$ boundaries. The role of bubble type solutions has
also been discussed by K.~Hashimoto in \cite{KojiBDecay}
especially from the viewpoint of time evolution of the tachyon
field.

\subsection{Bubble Embedding} \label{BubbleEmbedsubsec}

Motivated by the Witten bubble solution, we examine the following
analytic continuation of the embedding in the previous section. If
we make the analytic continuations $t \rightarrow \ i \chi$ and
$\theta \rightarrow \ \pi/2 - i \tau$ then the embedding functions
become
\begin{eqnarray} \label{bubbleembedding}
Z^0 & = & \chi \\
Z^1 & = & r sinh \tau \\
Z^i & = & r cosh \tau n^i \sp i=2,...,p\\
Z^{p+1} & = & \int \frac{B}{\sqrt{r^{2p-2} - r_0^{2p-2}}} dr \ .
\end{eqnarray}
The coordinates $Z^M$ for $M=0$ to $p$ are along the brane
worldvolume while $Z^{p+1}$ is again the direction of the
excitation of the configuration.  The embedding functions are all
real because we take this solution to be embedded into flat
Minkowski space with metric $\eta_{MN}=(1,-1,1,...,1)$.  The
induced metric on this brane is
\begin{equation} \label{bubblemetric}
ds^2_{bubble} = d\chi^2+\frac{1-A^2/r^{2p-2}}{1-(r_0/r)^{2p-2}}
dr^2 + r^2 (-d\tau^2+ cosh^2 \tau d\Omega_{p-2}^2)
\end{equation}
\begin{equation}
A^2 \equiv r_0^{2p-2}-B^2 \ .
\end{equation}
Here we impose the restriction $r_0^{2p-2} \geq B^2$ which is the
reverse inequality relative to the volcano solution; the
well-defined bubble solutions are analytic continuations of
volcano solutions with imaginary field strength. This solution has
an expanding bubble situated at the radius $r=r_0$ as will be
discussed shortly.  We also note that because
\cite{PenroseplanewaveGR} the embedding space of this bubble
metric has only one time, this solution is globally hyperbolic
which is just a way of saying that the spacetime has a Cauchy
surface. Having a target space with one embedding time
effectively means that each constant time slice is a Cauchy
surface.

One can ask why we did not drop the coordinate $\chi$ from the
solution and interpret this configuration as a lower dimensional
$p-1$ brane.  Although the coordinate $\chi$ plays no role in the
dynamics of the membrane embedding, one can not simply drop this
coordinate in general.  The reason why is that the general
solution has a magnetic field so dropping $\chi$ is equivalent to
the limiting case of no field strength.  Whereas the solution of
the previous section had a radial electric field, this solution
comes with a magnetic field in the ``angular directions'' but
which decreases radially
\begin{equation}
F_{\chi r}=  \frac{A}{\sqrt{r^{2p-2} - r_0^{2p-2}}} \ .
\end{equation}
Hence, although this solution has a string-like excitation, it can
not be interpreted as either a fundamental string or a D-string.

\subsection{Properties}

\begin{figure}[htb]
\begin{center}
\epsfxsize=6in\leavevmode\epsfbox{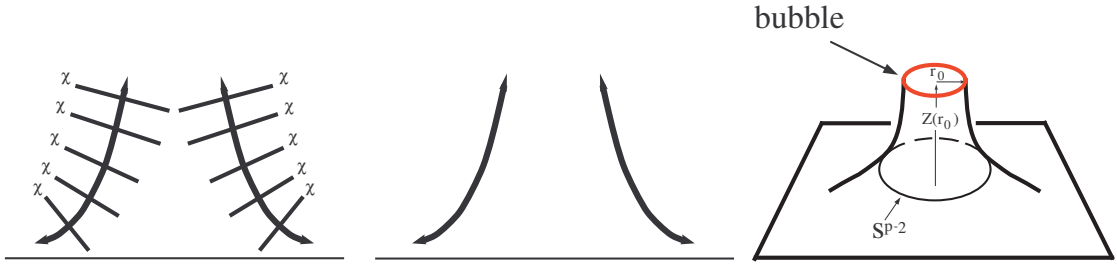} \caption{From left to
right we have: profile of bubble solution with transverse $\chi$
direction included, profile with $\chi$ suppressed, full view of
solution with $\chi$ suppressed.} \label{chi2}
\end{center}
\end{figure}

To obtain a clearer picture of the shape of this solution, see
also Figure~\ref{chi2}, consider the constant $\chi$ and
$Z^1=\tau=0$ slice of the $Dp$ brane bubble solution. We find
that this slice is the same as a constant time slice of a D(p-1)
volcano solution
\begin{equation}
ds^2_{bubble \ \tau=0} = \frac{1-A^2/r^{2p-2}}{1-(r_0/r)^{2p-2}}
dr^2 + r^2  d\Omega_{p-2}^2 \ .
\end{equation}
Therefore if we glue together two bubble solutions, we can also
interpret this configuration as a smooth brane and anti-brane
system. Just as in the case of the volcano solution we still have
the same change in orientation when we cross the horizon $r_0$.
This solution therefore is a time-dependent solution related to
brane and anti-brane annihilation.

\begin{figure}[htb]
\begin{center}
\epsfxsize=6in\leavevmode\epsfbox{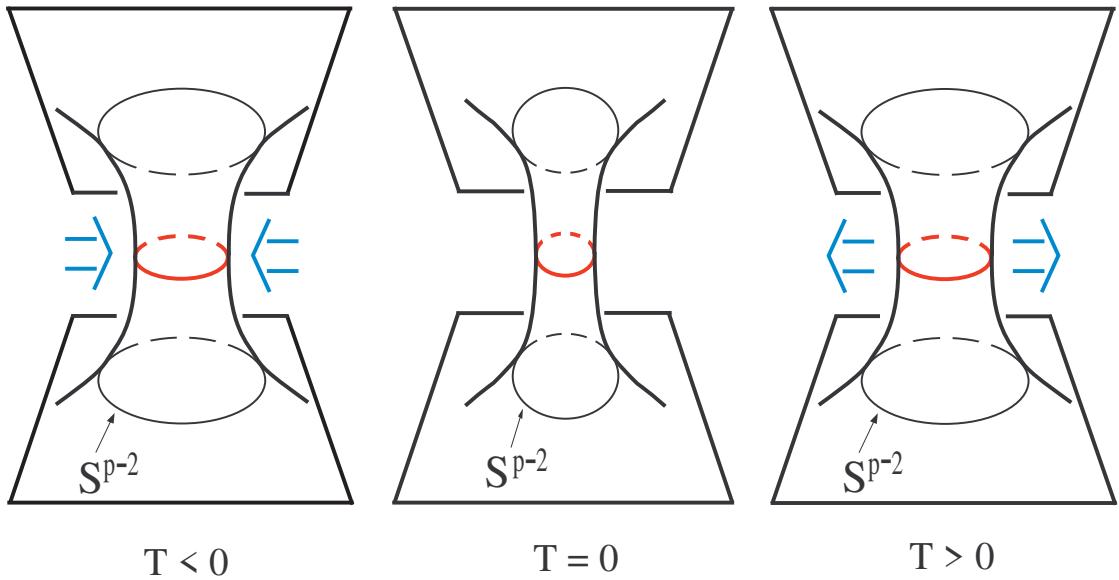} \caption{Bubble
solution for a p-brane configuration at three different times and
with the coordinate $\chi$ suppressed. The bubble is the minimum
volume $S^{p-2}$. The bubble itself undergoes contraction and
then expansion, while the volume of the brane expands and then
contracts.} \label{bubbletimeelapse}
\end{center}
\end{figure}

When there are no gauge fields, it is safe to drop the $\chi$
coordinate which does not play a role in this solution. In this
case it is possible to interpret the remaining solution as a
lower dimensional $p-1$ brane and it is easy to see why this
bubble solution is time dependent as compared to the volcano
solution.  Comparing the $p$ brane volcano solution and the $p-1$
dimensional bubble solution, we see that they have identical
excitation profiles.  If the $p$ brane volcano solution is just
massive enough to support this excitation, then the $p-1$ brane
can not be massive enough to keep this excitation static and
therefore the bubble expands.  In fact as shown in
Figure~\ref{bubbletimeelapse}, if we consider the full solution
it shows a brane and anti-brane rapidly coming in from infinity,
slowing to a stop when they become nearly parallel, and then
annihilating.  The energy of the branes is transferred into
accelerating the bubble wall. This solution has the relevant
boundary conditions to describe parallel brane and anti-brane
annihilation with no field strength.  This expanding bubble
solution derives its name from the fact that there is a minimal
volume sphere from the worldvolume perspective.

In general, examining the brane with the above worldvolume
coordinates leads to unusual behavior for the light cones
directions.  For example if we study the radial null geodesics, we
find that $(dr/d\tau)^2= r^2[1-(r_0/r)^{2p-2}]$ which does not go
to a fixed value when $r$ goes to infinity.  On the other hand, we
expect the spacetime to be asymptotically Minkowski since the
brane flattens out to $Z^{p+1}=0$.  Also it is unusual that there
is only a magnetic field in these coordinates since the spacetime
is time dependent.  This is a sign that to better understand the
dynamics, especially at infinity, we should express the system in
another set of coordinates to make contact with our usual
intuition.

\begin{figure}[htb]
\begin{center}
\epsfxsize=4in\leavevmode\epsfbox{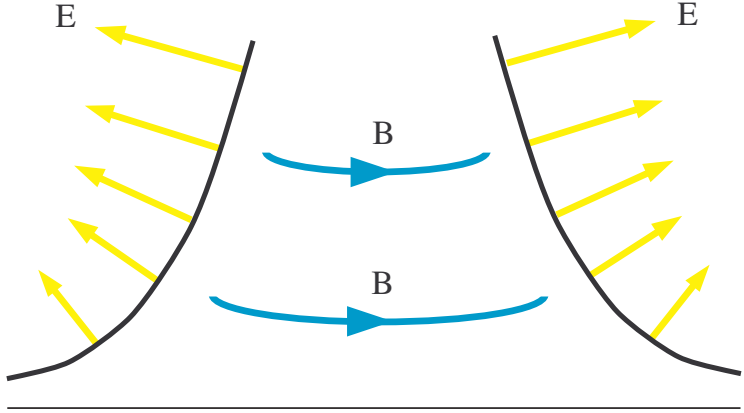} \caption{Side
profile of a bubble solution with the electric field along the
$\chi$ direction represented by arrows.} \label{Efield}
\end{center}
\end{figure}

One can choose the following coordinates to express the solution
\begin{equation}
W=r cosh \tau \sp T=r sinh \tau
\end{equation}
\begin{equation}
ds^2_{bubble}   =   d\chi^2 -dT^2 + dW^2 +
\frac{1-A^2/r_0^{2p-2}}{(\sqrt{W^2-T^2}/r_0)^{2p-2} -1}
\frac{(WdW-TdT)^2}{W^2-T^2} + W^2 d\Omega^2_{p-2}  \ .
\end{equation}
This is a natural choice of coordinates since the new time
coordinate along the brane, $T$, is the time coordinate of the
embedding space.  In these Cartesian coordinates the spacetime
becomes asymptotically Minkowski with the expected light cone
property $(dW/dT)^2 \rightarrow 1$ as $W$ goes to infinity. We
see then that the essence of the problem was that there was a
discrepancy between worldvolume time and embedding time.  In
these coordinates we have both an electric and a magnetic field
\begin{eqnarray}
F_{\chi T} & = & -\frac{T}{\sqrt{W^2-T^2}}
\frac{ A}{[(\sqrt{W^2-T^2})^{2p-2}-r_0^{2p-2}]^{1/2}} \\
F_{\chi W} & = & \frac{W}{\sqrt{W^2-T^2}} \frac{
A}{[(\sqrt{W^2-T^2})^{2p-2}-r_0^{2p-2}]^{1/2}} \ .
\end{eqnarray}
From the magnitudes of the field strength it appears that this
solution looks like a time dependent electric dipole and a
magnetic charge at large $W$ distances.  However the field
configurations have the electric field pointed along the $\chi$
direction and the magnetic field is in the angular directions. At
time $T=0$ there is only a magnetic field and no electric field.
If we stay at a fixed position $W$ which is far from the bubble,
$r_0$, then we see the electric field decreasing and the magnetic
field increasing. Another way to say this is that an initial
magnetic field gets pushed in the direction $\chi$, along which
the brane is smeared, in the form of an electric field. This
process is illustrated in figure~\ref{Efield}.  In this case the
brane and anti-brane annihilate in all spatial directions except
$\chi$ which remains unchanged for all times. This direction can
not be interpreted as a fundamental string since there is no
electric field at $T=0$.  It would be interesting however to see
if there are similar situations where $\chi$ can be identified as
a remnant of the annihilation process.

\begin{figure}[htb]
\begin{center}
\epsfxsize=6in\leavevmode\epsfbox{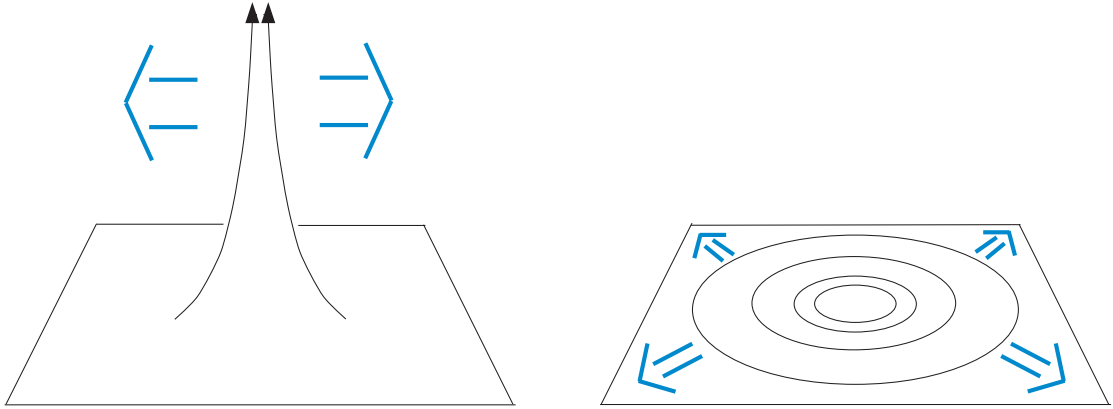} \caption{Time
evolution for the case of small bubble size compared with zero
bubble size.} \label{bubblelimit}
\end{center}
\end{figure}

The special case when $r_0$ approaches zero is also interesting.
As $r_0$ goes to zero, the excitation becomes narrower while the
distance between the brane and anti-brane increases; the width of
the throat increases in time just like for all the other bubble
solutions. However the case $r_0=0$ is qualitatively different
and has no excitation at all.  In this case we are describing only
a flat brane or an anti-brane.   A comparison of the solutions is
given in Figure~\ref{bubblelimit}. The configuration is flat
since the constraint $A^2+B^2=r_0^{2p-2}$, implies if $r_0=0$
then $A$ and $B$ are zero.  In this case we can drop the $\chi$
direction and the solution is
\begin{eqnarray} \label{deSitterembedding}
Z^1 & = & r sinh \tau \\
Z^i & = & r cosh \tau n^i \sp i=2,...,p\\
ds^2_{limit} & = & dr^2 + r^2 (-d\tau^2+ cosh^2 \tau
d\Omega_{p-2}^2)
\end{eqnarray}
which is just Minkowski space in unusual coordinates. Setting $r$
constant, which is de Sitter space, is not a solution.

\subsection{Comparison to Bubble of Nothing}

Here we further examine the special case with no field strength.
The induced metric on this non-BPS solution is
\begin{equation} \label{DBIbubblemetric}
ds^2_{bubble} = d\chi^2+\frac{ dr^2}{1-(r_0/r)^{2p-2}} + r^2
(-d\tau^2+ cosh^2 \tau d\Omega_{p-2}^2)
\end{equation}
where for comparison purposes we have reverted back to the
spherical coordinate system in Eq.~\ref{bubbleembedding}.  This is
quite similar to the Witten bubble metric
\begin{equation} \label{Wittenbubblemetric}
ds^2_{WB} = [1- (\frac{r_0}{r})^{D-2}] d\chi^2 + \frac{
dr^2}{1-(r_0/r)^{D-2}} + r^2 (-d\tau^2+ cosh^2 \tau
d\Omega_{D-2}^2)
\end{equation}
which is a vacuum gravity solution for $D+1$ spacetime dimensions
with $r \geq r_0$.  To make the Witten bubble metric regular at
$r=r_0$, we must impose the periodicity condition $\chi=\chi + 4
\pi r_0/(D-2)$.  Asymptotically the Witten bubble is flat space
times a circle whose radius is proportional to the bubble size.
Although this periodicity condition is apparently not required
for the DBI solution (since $\chi$ is independent of the other
coordinates), actually the periodicity condition is reflected in
the fact that we had to join together the two halves of a bubble
solution at $r_0$ to form a smooth wormhole.  In other words the
distance between the brane and anti-brane is fixed by the minimum
size bubble due to the tension of the branes; the only way to
change this relationship is by adding charge to the branes.

Although these two solutions are similar, there is never a case
where they exactly coincide even if we neglect the $\chi$
direction. Setting $D=p$ we find that the DBI solution changes
much more quickly in the radial direction than the Witten
bubble.  One can view a slice of the Witten bubble solution in
$D=2p$ as being exactly the bubble solution. In this case the
time coordinate is embedded correctly so dynamics on the brane
configuration really are the same as that of a slice of the
Witten bubble.  Comparing the whole solutions, however, the new
bubble solution is closer to a $Z_2$ projection
\cite{FabingerHorava} along the circle direction of the Witten
bubble. It is interesting that the DBI solutions are similar only
to the Witten bubbles in odd spacetime dimensions. One wonders if
this is a reflection of the different causal structures of branes
of even versus odd dimension\cite{GibbonsHorTownResolveSing}.  If
so then it suggests the existence of new DBI solutions.

The Witten bubble has also been called a bubble of nothing since
there is a ``hole'' in space due to the fact that we always stay
in the region $r \geq r_0$; naively being inside the hole would
give a spacetime with three timelike directions.  One might ask
if spacetime becomes singular or if there is a boundary at $r_0$.
Given that the new bubble in Eq.~\ref{bubblemetric} is smooth and
boundary free, and given the similarities of the two bubble
solutions, it should not be surprising that the Witten bubble is
smooth and boundary free. To better understand the spacetime near
$r_0$ and see that there is no singularity there, it is
convenient to make the coordinate transformation
$y=\sqrt{1-(r_0/r)^{D-2}}$. This region is undergoing inflation
in the radial direction as can be seen by Taylor expanding the
metric
\begin{equation} \label{WBmetricexpansion}
ds^2_{WB} \approx  r_0^2 [ (1 + \frac{2}{D-2}y^2) (-d\tau^2 +
cosh^2 \tau d\Omega^2_{D-2} ) + (\frac{2}{D-2})^2(1 +
2\frac{D-1}{D-2}y^2)dy^2 ] \ .
\end{equation}
So near the bubble, and actually at all fixed radius points, the
spacetime first undergoes smooth polynomial expansion and then
exponential inflation at late times.  The bubble accelerates from
zero velocity to the speed of light at late times. The same
conclusion also holds for the bubble solution of
Eq.~\ref{bubbleembedding} which also undergoes inflation along
the worldvolume directions. A quick way to see this is examine
the constraint $-(Z^1)^2 + \sum_i (Z^i)^2= r^2$.  This
constraint, along with the fact that the brane flattens out at
infinity, shows that the bubble approaches the speed of light at
late times.  Finally, we note that of the two bubbles, the Witten
bubble accelerates from zero velocity to the speed of light more
slowly for $p=D$.

The surface of the Witten bubble undergoes contraction and then
expansion.  In fact, this surface of the bubble in
Eq.~\ref{WBmetricexpansion} has the metric of de Sitter space
\begin{equation}
r_0^2 (-d\tau^2+ cosh^2 \tau d\Omega_{D-2}^2) \ .
\end{equation}
The bubble solution is in effect a way to embed de Sitter space
into an asymptotically flat background solution.
Ref.~\cite{cleanbubbles} gave a discussion as to whether one can
regard the Witten bubble to be a full solution of classical
string theory.  Having obtained a similar bubble solution one can
ask the same question, do we have a string theory solution.
Because this embedding satisfies the DBI equations, this solution
naively captures the $\alpha^\prime$ contributions in the small
string coupling limit. However at our level of approximation we
are neglecting possible higher order derivative corrections and
the tachyon field. Since these solutions are not supersymmetric
one would question if, by neglecting the tachyon field, one is
being consistent.  Although we have found new solutions to the
DBI equations of motion, it is not apparent if these are also
solutions of string theory.

Further properties of the Witten bubble can be found in
\cite{Wittenbubble, FabingerHorava, cleanbubbles}.

\subsection{Low Dimensional Bubbles}

In analyzing the rest of the lower dimensional bubbles, we drop
the $\chi$ direction.  Then for the two brane case, $p=3$, we find
that the profile of the excitation $Z^4$ is not logarithmic and
flattens out at infinity. This bubble solution should be relevant
to the decay of a parallel two-brane and anti-brane pair.  In
contrast, the volcano solutions for the case of two-branes had
logarithmic behavior at infinity.  Although the volcano solution
is a particular two brane configuration, it is not related to
annihilation of parallel two-branes.  When $p=3$ we can also find
another time dependent solution. This solution has the same
spatial embedding as the other bubbles except it has an electric
field in the angular $\phi$ direction
\begin{equation}
F_{\tau \phi}= \sqrt{-B^2+r_0^4} cosh \tau
\end{equation}
which comes from analytically continuing the volcano solution
with a magnetic field in Eq.~\ref{volcanomagnets}.   In this case
the $\chi$ direction does not seem to play a role and so could be
dropped. We can therefore regard this too as a time dependent
two-brane configuration.

For the case $p=2$, we find a one dimensional time dependent
solution which one can explicitly write down. When $p=1$ we have
a point particle and since there is no angular coordinate to
analytically continue, it might appear that this configuration
has no time evolution.  Realizing, however, that we can also
analytically continue the Cartesian coordinates, we find solutions
for a free particle.   A discussion of the annihilation of
zero-branes can be found in \cite{KojiBDecay}.

\newsection{Spacelike Brane} \label{SbraneSec}

In this section we find further solutions of the DBI equations of
motion.  These solutions will be very different from the bubble
and volcano solutions we have discussed so far and in fact they
will be embedded in spaces with two times.  Yet even with the
appearance of two times, their worldvolume will have just one
timelike direction and they will share similarities with known
supergravity S-brane solutions
\cite{PopeandthreeCosmologicalStringTheory,
PopeandtwoBranesandCosmology, GutperleStrominger, ChenGutperle,
MyersSbrane}.

\subsection{S-brane embedding}
Analytically continuing the embedding of the volcano solution in
Eq.~\ref{stringembedding} by taking $t \rightarrow \ i \chi$,
$\theta \rightarrow \ -i \theta$ and $r\rightarrow \ i \tau$ we
obtain
\begin{eqnarray}
Z^0 & = & \chi \\
Z^1 & = & \tau cosh \theta \\
Z^i & = & \tau sinh \theta n^i \sp i=2,...,p\\
Z^{p+1} & = & \int \frac{B}{\sqrt{\tau^{2p-2} - \tau_0^{2p-2}}}
d\tau  \ .
\end{eqnarray}
Since we made a triple analytic continuation the embedding space
is not Minkowski and its metric $G_{MN}=(1,-1,1,...,1,-1)$ has two
times.  The induced metric on this solution is
\begin{equation}
ds^2= d\chi^2 -\frac{1+A^2/\tau^{2p-2}}{1-(\tau_0/\tau)^{2p-2}
}d\tau^2 + \tau^2[ d\theta^2 + sinh^2\theta d\Omega^2_{p-2}]
\end{equation}
\begin{equation}
A^2=B^2-\tau_0^{2p-2}
\end{equation}
and the field strength
\begin{equation}
F_{\chi \tau}=  \frac{-A}{\sqrt{\tau^{2p-2} - \tau_0^{2p-2}}}
\end{equation}
is an explicitly time dependent electric field.  Here we must take
$B^2\geq \tau_0^{2p-2}$ just like for the volcano solution. Even
though the embedding space has two times, this S-brane type
solution is still well behaved in two respects.  The DBI action
is still real and it is possible to discuss time-evolution on the
brane. The induced metric on this solution has at most only one
independent time-like direction and if we stay in region $\tau
\geq \tau_0$ then $\tau$ can be used to parametrize the
worldvolume time.  If we naively try to take $\tau \leq \tau_0$
then the worldvolume is spacelike.  In this paper we will not
discuss this spacelike region, or attempt to discuss any
relationship between the two embedding times and the possible
holographic reconstruction of a timelike direction proposed for
spacelike branes.

For the case without charge, $A=0$, the induced metric
\begin{equation}
ds^2= d\chi^2 -\frac{d\tau^2}{1-(\tau_0/\tau)^{2p-2} } + \tau^2[
d\theta^2 + sinh^2\theta d\Omega^2_{p-2}]
\end{equation}
bears some similarity to the known S-brane supergravity solutions
in $p+1$ dimensions.  For example the metric of an S0-brane
solution in four dimensions discussed in
\cite{GutperleStrominger} is
\begin{equation}
ds^2_{S0}= \frac{\tau_0^2}{Q^2}(1-(\tau_0/\tau)^{2}) d\chi^2
-\frac{Q^2}{\tau_0^2}\frac{d\tau^2}{1-(\tau_0/\tau)^{2} } +
\frac{Q^2}{\tau_0^2}  \tau^2[ d\theta^2 + sinh^2\theta d\phi^2] \
.
\end{equation}
However due to the fact that the known S-branes contain a dilaton,
a form field, or a field strength, there is actually another
solution to Einstein's equations which is more like the DBI
solution above. It is the triple analytic continuation of a $D+1$
Schwarzschild black hole
\begin{equation}
ds^2= (1-(\tau_0/\tau)^{D-2}) d\chi^2 -
\frac{d\tau^2}{1-(\tau_0/\tau)^{D-2} } + \tau^2[ d\theta^2 +
sinh^2\theta d\Omega^2_{D-2}]
\end{equation}
where we have taken the analytic continuation $t \rightarrow i
\chi$, $r\rightarrow \ i \tau$ and $\theta\rightarrow\ i \theta$
of the Schwarzschild solution.  Since this metric is nothing but
the analytic continuation of the Schwarzschild metric, it clearly
satisfies the vacuum Einstein equations.

\subsection{Properties}

As mentioned in Subsection~\ref{BubbleEmbedsubsec}, if the
embedding space had only one time then this would imply that the
embedded solution is globally hyperbolic. Since it turned out
that our embedding space has two times, one then wonders if this
was just an artifact of the embedding or if it not possible to
embed this solution into a space with one time. If not, then
there are some possible reasons why one time-like direction does
not suffice. The first reason is that this spacelike brane seems
to have a singularity at $\tau_0$. This induced metric has
singular components
when $\tau=\tau_0$.   It is known that singularities in the time
direction can lead to spaces not being globally hyperbolic. The
second reason is that spacelike branes are supposed to represent
a finely tuned configuration for a system. Perhaps this fine
tuning could mean that the system is correlated so it is not
enough to specify the initial conditions of the system at one
moment in time so there is no Cauchy surface. The size of the
correlation effects could then give the S-branes a finite width.

This new solution does not appear to be a brane and anti-brane
configuration but instead seems to be an unstable brane
configuration.  Let us use the worldvolume time to analyze the
shape of this brane. It is possible to use the parameter $\tau$
to obtain sensible one time evolution along the worldvolume as
long as we take $\tau \geq \tau_0$.  In the above spherical
coordinate system we find some interesting features of this
solution.  For comparison we first review the bubble case where
the spatial directions satisfied the relationship $(Z^i)^2=r^2
cosh^2\tau$. This shows us that at time $\tau=0$ the solution
starts at rest and that at late times the solution has
accelerated to the speed of light. For the S-brane type solution
we have $(Z^i)^2=\tau^2 sinh^2 \theta$. Yet since we also take
$\tau\geq \tau_0$, the solution  begins with some finite initial
velocity proportional to $\tau_0$. Further there is a gradient of
initial velocities as we vary $\theta$. The spatial origin which
is located at $\theta=0$ does not move while the points at large
$\theta$ move very quickly.

\begin{figure}[htb]
\begin{center}
\epsfxsize=6in\leavevmode\epsfbox{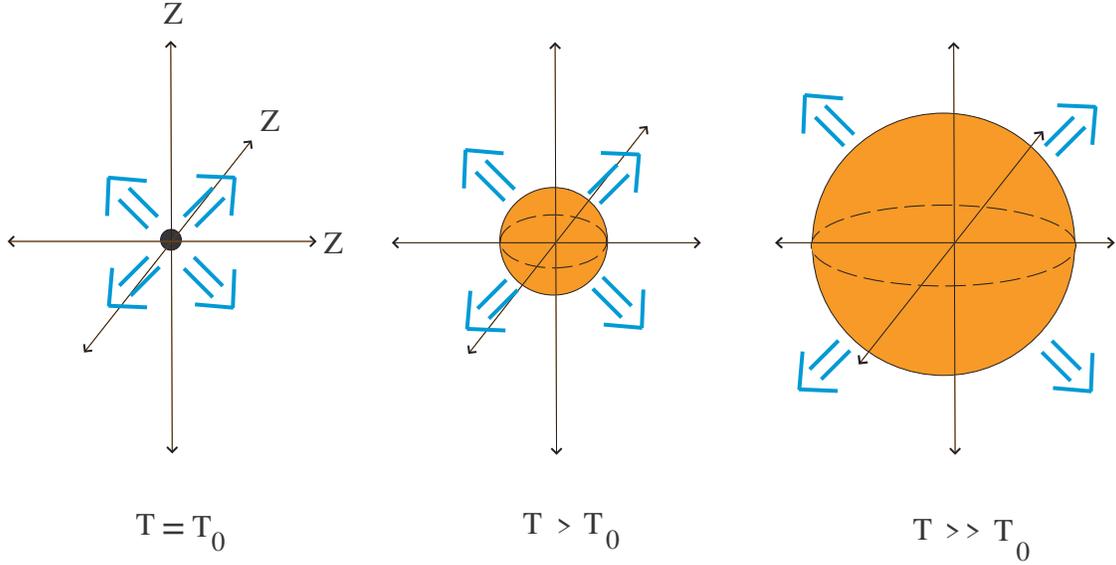} \caption{Along the
spatial directions, the S-brane type solution begins at a point
and expands outwards.} \label{bangpic}
\end{center}
\end{figure}

\begin{figure}[htb]
\begin{center}
\epsfxsize=6in\leavevmode\epsfbox{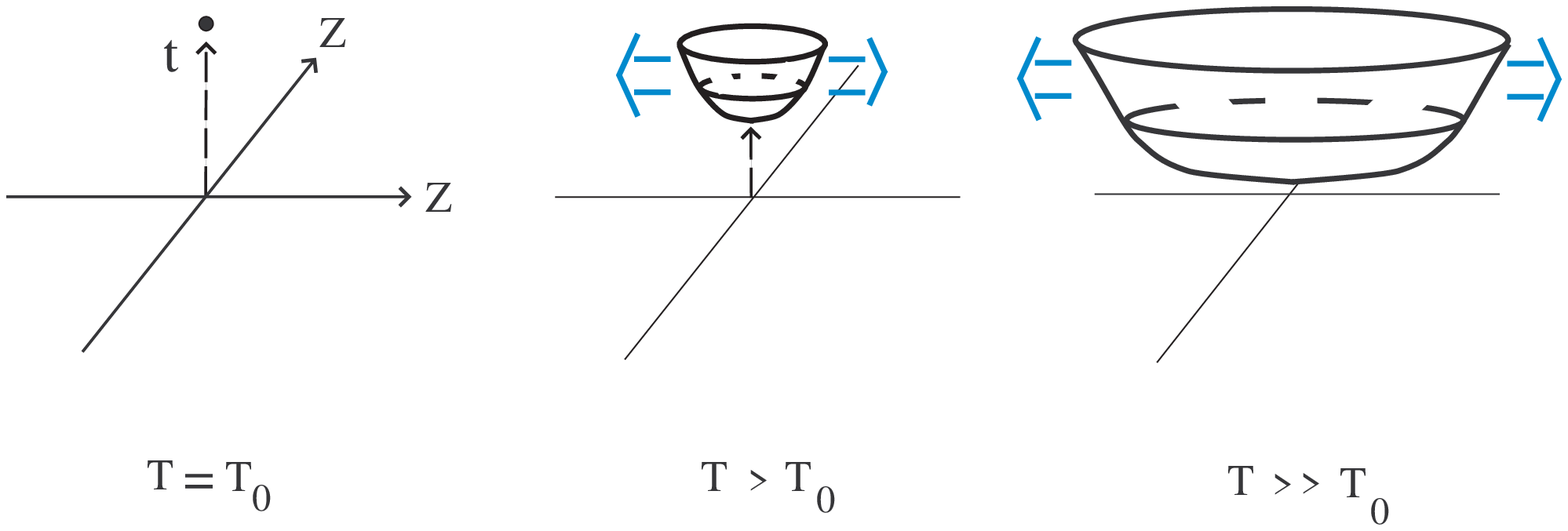} \caption{In the
full embedded space, the S-brane type solution begins highly
peaked in a timelike direction and relaxes into a flat
configuration.} \label{Sbranepic}
\end{center}
\end{figure}

It is also instructive to analyze the time evolution of this
brane using the coordinate system
\begin{equation}
T=\tau cosh \theta, \hsp{.3} W=\tau sinh \theta
\end{equation}
which is similar to the coordinate transformation for the
bubble.  In this case the induced metric becomes
\begin{equation}
ds^2_{S-type}   =   d\chi^2 - dT^2 + dW^2 -
\frac{1+A^2/\tau_0^{2p-2}}{(\sqrt{T^2-W^2}/\tau_0)^{2p-2} -1}
\frac{(WdW-TdT)^2}{T^2-W^2} + W^2 d\Omega^2_{p-2} \ .
\end{equation}
In terms of the $W,T$ coordinates, the restriction $\tau^2 \geq
\tau_0^2$ becomes
\begin{equation}
W^2\leq T^2 - \tau_0^2
\end{equation}
so we live in the ball bounded by a maximum sphere.  Also the
metric singularity at $\tau=\tau_0$ turns into a singularity when
$W^2=T^2-\tau_0^2$. Along the spatial directions, $Z^i$ with
$i=2,...,p$, this brane can be said to represent an expanding (or
contracting) brane with the condition that the hole always begins
with zero size since $\tau$ and hence $T$ starts at $\tau_0$. The
time evolution is shown schematically in Figure~\ref{bangpic}.
Although the brane does not start off with an infinite velocity,
it expands to infinite size. At late times in this coordinate
system, the brane becomes essentially flat Minkowski space. If we
include the excitation direction, this brane has the form of a
peaked excitation which eventually flattens out as shown in
Figure~\ref{Sbranepic}.  In the two previous cases, it was shown
that we could glue together solutions to form a brane and
anti-brane system.  Although in this case since the excitation
direction is timelike, since there is only one worldvolume time
it seems possible also to perform the gluing procedure.  From the
perspective of the brane observers the excitation direction
serves only to time dilate events.  The worldvolume then becomes
topologically a double cover of a ball and instead of going to the
edge of the ball we arrive on an identical second surface.

\subsection{Low Dimensional Solutions}

For the case $p=3$, we can find another spacelike brane with time
dependent electric field
\begin{equation}
F_{\theta \phi}= \sqrt{B^2-\tau_0^4} sinh \theta  .
\end{equation}
This is the analytic continuation of the volcano solution with a
magnetic field.  When $\tau_0=0$ the solution is non-trivial.

In treating the other solutions of low dimensionality, take
$\chi=0$ and $\tau > \tau_0$.  For the case $p=2$, take a constant
timelike slice of $\tau$. Then in this case $(Z^1)^2-(Z^2)^2$ is
constant and the span of $\theta$ gives half of a hyperbola at a
constant $Z^3$ height. As we evolve to more positive $\tau$
times, the brane moves away from the origin and decreases in
$Z^3$ height.  Due to the low dimensionality of this solution it
is also possible to reinterpret the embedding space with metric
$(-1,1,-1)$ as having only one time. This solution has the same
form as the time dependent $p=2$ bubble solution. When $p$ is
greater than two, then it is not possible to reinterpret the
embedding space as having one time.

\begin{figure}[htb]
\begin{center}
\epsfxsize=2in\leavevmode\epsfbox{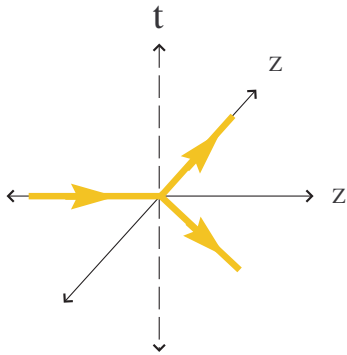}
\caption{Tachyon interaction process} \label{vertex}
\end{center}
\end{figure}

For the case $p=1$, the worldvolume is zero dimensional, and we
are describing a point like object. We find a freely moving
object which is purely in the time direction since the embedding
space has metric $(-1,-1)$. However we can again interpret this
as a purely spatial trajectory and add an extra embedding
dimension $t$ of plus signature.  The embedding space now has the
metric $(-1,-1,1)$. In this case the dynamics of this object are
equivalent to those of a freely propagating tachyon. These
spacelike branes are tachyons with spacelike trajectories which
exist only at one moment of time $t$ in the embedding space. It
is interesting to explore to what extent string junctions can be
interpreted as a tachyon interaction process as shown in
Figure~\ref{vertex}.

Further properties of these solutions will be discussed elsewhere.

\newsection{Conclusions}

We have provided time dependent solutions to the DBI equation of
motions.  These solutions are similar to Witten's bubble of
nothing, de Sitter space and to S-branes and one is tempted to
consider them as ``dual'' of the closed string gravity
solutions.  The interpretation of the S-brane type solution is
not very clear although it seems to be a generalization of a
tachyon.  It is an unstable brane configuration with
singularities, and its dynamics take place in a background space
with two times.  Although the interpretation of two times is not
fully understood, the worldvolume has only one timelike direction
which we used for time evolution.

The bubble solution has a clear interpretation as being a time
dependent brane and anti-brane annihilation process that can be
solved exactly and which resembles some cosmological scenarios.
For example one could imagine that we live in an inflationary
universe due to the annihilation of a $D3-\overline{D3}$ pair.  A
problem immediately arises since this space has eternal
inflation. However since more than one bubble can nucleate, the
interactions of different bubbles could potentially serve as a
qualitative reason why inflation should end.  Another problem
with a such a scenario however is how one could account for the
fact that such branes actually got close enough for a wormhole to
have a reasonable chance to form.  Taking into account the mutual
attraction between the branes could potentially solve this
problem.

To increase the chances for wormholes to nucleate, one could
increase the dimensionality of the branes.  For example we could
take a $D4-\overline{D4}$ pair which is much more likely to
intersect. If the branes intersect then they will be close enough
in general for wormholes to form with non-negligible
probability.  In this case one wonders if it is possible to have
our observable universe located on the bubble which is four
dimensional de Sitter space with matter accumulating near the
bubble as it sweeps throughout the brane.  One can also examine
the case of fivebranes \cite{StephonDDbarInflation} and look for
a new decay process.  In this case one would look for a vortex
solution which would be our three brane universe.

In this paper we have discussed the open string tachyon from the
DBI point of view.  Of course one wonders about the closed string
tachyons.  In general it appears to be a difficult question to
imagine how to give a formulation where both instabilities are
treated concurrently. It is possible however to make some
qualitative statements about known instabilities and see if they
agree with expectations. For example in a set up similar to
Fabinger and Horava but for p-branes instead of $E_8$ boundaries,
one can calculate that the orbifold projection of the Witten
bubble has a nucleation probability proportional to
$exp(-L^{p-2}/g^2 l_s^{p-2})$ where $L$ is the separation
distance between branes. In the case of the DBI bubble we find
that the nucleation rate is proportional to $exp(-L^{p+1}/g
l_s^{p+1})$, which has the same form as the decay rate calculated
\cite{KojiBDecay} using the tachyonic DBI action.  In the case of
very large separation the Witten bubble is the more dominant
decay process. In the case where the brane separation is of the
order of the string scale, the Witten bubble dominates for large
string coupling while the DBI bubble dominates for small string
coupling.  The expectation is that for light branes, the
backreaction is small and so the open string tachyon should
dominate, while for heavy branes the backreaction is large and so
the closed string tachyon should dominate.

For small string coupling the above results seem to lead to the
following possible picture of the tachyon decay process for a
brane and anti-brane. A brane and anti-brane at large separation
create Witten bubbles between them and approach each other. The
bulk of spacetime is annihilating. When the brane separation
approaches the string scale, the other bubbles take over and we
proceed with the safer decay process where just the brane and
anti-brane annihilate. Although both tachyons are present, the
safer one takes over and saves the bulk of the universe while
annihilating the branes. In this account we are neglecting the
interaction of the various bubbles which is expected to be
complicated.  An explicit solution describing two Witten bubbles
has recently been found \cite{Horowitz2bubbles}.

Given the above qualitative statements, one can also attempt to
make some contact with the homogenous tachyon decay process
discussed by Sen.  Although the bubble decay process is
inhomogeneous, in the limit of small string coupling and brane
separation, the probability of bubble production becomes very
large. In general one would then expect to see an almost uniform
distribution of bubble type solutions when the branes are nearly
coincident.

Finally it would be interesting to see if solutions with time
dependent field strength can give a physical picture for the
problem of the U(1) gauge field \cite{IIBornotIIB,
WittenDbranesKtheory, SennonBPSaction, SenUniversalTachyon,
HoriConfinementBrane, HoriStringFluid}.  New explicit time
dependent solutions describing the fate of the electric field can
be found in \cite{SenDbraneEfield, KojiGYCarrolian} and will also
play a role in a future paper.

\noindent {\bf Acknowledgements}

\noindent I would like to express my gratitude to the following
people for stimulating and invaluable discussions: Chiang-Mei
Chen, Sergey Cherkis, Koji Hashimoto, Pei-Ming Ho, Matt Kleban,
Ashoke Sen, Hyun Seok Yang and Hoi-Lai Yu.

I am supported by the Academia Sinica and would like to thank its
members.  I would also like to thank Pei-Ming Ho, the COSPA
project and NCTS for additional support.

\bibliographystyle{ieeetr}
\bibliography{bibliography}

\begin{thebibliography}{10}

\bibitem{SenRollingTachyon}
{Ashoke Sen}, ``{Rolling Tachyon},'' {\em JHEP}, hep-th/0203211.

\bibitem{SenTachyonMatter}
{Ashoke Sen}, ``{Tachyon Matter},'' hep-th/0203265.

\bibitem{HullTimeTdual}
{C. M. Hull}, ``{Timelike T-duality, de Sitter space, Large N Gauge Theories
  and Topological Field Theories},'' {\em JHEP}, vol.~{\bf{9807}}, pp.~021,
  1998, hep-th/9806146.

\bibitem{HullldeSitterMtheory}
{C. M. Hull}, ``{De Sitter Space in Supergravity and M Theory},'' {\em JHEP},
  vol.~{\bf{0111}}, pp.~012, 2001, hep-th/0109213.

\bibitem{TyeBraneInteractInflation}
{Nicholas Jones, Horace Stoica and S.-H. Henry Tye}, ``{Brane Interaction as
  the Origin of Inflation},'' {\em JHEP}, vol.~{\bf{9812}}, pp.~019, 1998,
  hep-th/9810188.

\bibitem{DvaliTyeBraneInflation}
{G. Dvali and S. -H. Henry Tye}, ``{Brane Inflation},'' {\em Phys. Lett.},
  vol.~{\bf{B450}}, pp.~72, 1999, hep-th/9812483.

\bibitem{DvaliShafiDbraneInflation}
{G. Dvali, Q. Shafi and S. Solganik}, ``{D-brane Inflation},'' hep-th/0105203.

\bibitem{StephonDDbarInflation}
{Stephon H. S. Alexander}, ``{Inflation from D-Dbar Annihilation},'' {\em Phys.
  Rev. D.}, vol.~{\bf{65}}, pp.~023507, 2002, hep-th/0105032.

\bibitem{BurgessInflationBraneaBraneUniverse}
{C. P. Burgess, M. Majumdar, D. Nolte, F. Quevedo, G. Rajesh and R. -J. Zhang},
  ``{The Inflationary Brane-Antibrane Universe},'' {\em JHEP},
  vol.~{\bf{0107}}, pp.~047, 2001, hep-th/0105204.

\bibitem{GibbonsRollTachyon}
{G. W. Gibbons}, ``{Cosmological Evolution of the Rolling Tachyon},'' {\em
  Phys. Lett.}, vol.~{\bf{B537}}, pp.~1--4, 2002, hep-th/0204008.

\bibitem{PopeandthreeCosmologicalStringTheory}
{H. Lu, S. Mukherji, C.N. Pope and K.-W. Xu}, ``{Cosmological solutions in
  string theories},'' {\em Phys. Rev.}, vol.~{\bf{D55}}, pp.~7926--7935, 1997,
  hep-th/9610107.

\bibitem{GutperleStrominger}
{Michael Gutperle and Andrew Strominger}, ``{Space-like Branes},'' {\em JHEP},
  vol.~{\bf{0204}}, pp.~018, 2002, hep-th/0202210.

\bibitem{PopeandtwoBranesandCosmology}
{H. Lu, S. Mukherji and C.N. Pope}, ``{From p-branes to cosmology},'' {\em
  Int.J.Mod.Phys.}, vol.~{\bf{A14}}, pp.~4121--4142, 1999, hep-th/9612224.

\bibitem{ChenGutperle}
{Chiang-Mei Chen, Dmitri V. Gal'tsov and Michael Gutperle}, ``{S Brane
  Solutions in Supergravity Theories},'' hep-th/0204071.

\bibitem{MyersSbrane}
{Martin Kruczenski, Robert C. Myers and Amanda W. Peet}, ``{Supergravity
  S-Branes},'' {\em JHEP}, vol.~{\bf{0205}}, pp.~039, 2002, hep-th/0204144.

\bibitem{DegerKayaSbrane}
{N. S. Deger and A. Kaya}, ``{Intersecting S-Brane Solutions of D=11
  Supergravity},'' hep-th/0206057.

\bibitem{CallanM}
{Curtis G. Callan, Jr. and J. M. Maldacena}, ``{Brane Dynamics From the
  Born-Infeld Action},'' {\em Nucl. Phys. B}, vol.~{\bf{513}}, pp.~198--212,
  hep-th/9708147.

\bibitem{GibbonsBion}
G.~W. Gibbons, ``{Born-Infeld particles and Dirichelet p-branes},'' {\em Nucl.
  Phys. B}, vol.~{\bf{514}}, pp.~603--639, hep-th/9709027.

\bibitem{Wittenbubble}
{Edward Witten}, ``{Instability of the Kaluza-Klein Vacuum},'' {\em Nuc. Phys.
  B}, vol.~{\bf{195}}, pp.~481--492, 1982.

\bibitem{Giveonreview}
{A. Giveon and D. Kutasov}, ``{Brane Dynamics and Gauge Theories},'' {\em Rev.
  Mod. Phys.}, vol.~{\bf{71}}, pp.~983--1084, 1999, hep-th/9802067.

\bibitem{FabingerHorava}
{Michal Fabinger and Petr Horava}, ``{Casimir Effect between World Branes in
  Heterotic M Theory},'' {\em Nucl. Phys. B}, vol.~{\bf{580}}, pp.~243--263,
  2000, hep-th/0002073.

\bibitem{KojiBDecay}
{Koji Hashimoto}, ``{Dynamical Decay of Brane, anti-Brane and Dielectric
  Brane},'' hep-th/0204203.

\bibitem{PenroseplanewaveGR}
{R. Penrose}, ``{A remarkable property of plane waves in General Relativity},''
  {\em Rev. Mod. Phys.}, vol.~{\bf{37 N. 1}}, p.~215, 1965.

\bibitem{GibbonsHorTownResolveSing}
{G. W. Gibbons, G. T. Horowitz and P. K. Townsend}, ``{Higher-dimensional
  Resolution of Dilatonic Black-hole Singularities},'' {\em Class. Quantum
  Grav.}, vol.~{\bf{12}}, pp.~297, 1995, hep-th/9410073.

\bibitem{cleanbubbles}
{Ofer Aharony, Michal Fabinger, Gary T. Horowitz and Eva Silverstein}, ``{Clean
  Time Dependent String Backgrounds from Bubble Baths},'' hep-th/0204158.

\bibitem{Horowitz2bubbles}
{Gary T. Horowitz, Kengo Maeda}, ``{Colliding Kaluza-Klein Bubbles},''
  hep-th/0207270.

\bibitem{IIBornotIIB}
{M. Srednicki}, ``{IIB or not IIB},'' {\em JHEP}, vol.~{\bf{9808}}, pp.~005,
  1998, hep-th/9807138.

\bibitem{WittenDbranesKtheory}
{E. Witten}, ``{D-Branes and K-Theory},'' {\em JHEP}, vol.~{\bf{9812}},
  pp.~019, 1998, hep-th/9810188.

\bibitem{SennonBPSaction}
{Ashoke Sen}, ``{Supersymmetric World-volume Action for Non-BPS D-branes},''
  {\em JHEP}, vol.~{\bf{9910}}, pp.~008, 1999, hep-th/9909062.

\bibitem{SenUniversalTachyon}
{Ashoke Sen}, ``{Universality of the Tachyon Potential},'' {\em JHEP},
  vol.~{\bf{9912}}, pp.~027, 1999, hep-th/9911116.

\bibitem{HoriConfinementBrane}
{O. Bergman, K. Hori and Piljin Yi}, ``{Confinement on the Brane},'' {\em Nucl.
  Phys.}, vol.~{\bf{B580}}, pp.~289, 2000, hep-th/0002223.

\bibitem{HoriStringFluid}
{Gary Gibbons, Kentaro Hori and Piljin Yi}, ``{String Fluid from Unstable
  D-branes},'' {\em Nucl. Phys. B}, vol.~{\bf{596}}, pp.~136--150, 2001,
  hep-th/00009061.

\bibitem{SenDbraneEfield}
{Partha Mukhopadhyay and Ashoke Sen}, ``{Decay of Unstable D-branes with
  Electric Field},'' hep-th/0208142.

\bibitem{KojiGYCarrolian}
{Gary Gibbons, Koji Hashimoto and Piljin Yi}, ``{Tachyon Condensates,
  Carrollian Contraction of Lorentz Group, and Fundamental Strings},''
  hep-th/0209034.

\end{thebibliography}


\end{document}